\documentclass[12pt]{article}
\usepackage{amssymb}
\def\lddots{\mathinner{\mkern1mu\raise1pt\hbox{.}\mkern2mu
\raise4pt\hbox{.}\mkern2mu\raise7pt\vbox{\kern7pt\hbox{.}}\mkern1mu}}
\makeatletter
\def\numberbysection{\@addtoreset{equation}{section}
\def\theequation{\thesection.\arabic{equation}}}
\makeatother

\usepackage{epic}
\usepackage{eepic}
\usepackage{graphicx}
\usepackage{latexsym}
\usepackage{amssymb}
\numberbysection

\newcommand{\be}{\begin{eqnarray}}
\newcommand{\ee}{\end{eqnarray}}
\newcommand{\non}{\nonumber}

\newcommand{\tr}{\mathop{\rm tr}\nolimits}

\newcommand{\pt}{\hat{p}}
\newcommand{\qt}{\hat{q}}

\begin{document}

\begin{titlepage}
\strut\hfill
%
\begin{center}
{\bf {\Large A note on the boundary spin $s$ XXZ chain}}
\\[0.5in]

\vskip 0.2cm

{\large Anastasia Doikou\footnote{e-mail: doikou@bo.infn.it}}

\vspace{10mm}

{\small \emph{University of Bologna, Physics Department, INFN Section \\
Via Irnerio 46, 40126 Bologna, Italy}}

\end{center}
%

\vfill

\begin{abstract}

The open spin $s$ XXZ model with non-diagonal boundaries is
considered. Within the algebraic Bethe ansatz framework and in the
spirit of earlier works we derive suitable reference states. The
derivation of the reference state is the crucial point in this
investigation, and it involves the solution of sets of difference
equations. For the spin $s$ representation, expressed in terms of
difference operators, the pseudo-vacuum is identified in terms of
$q$-hypergeometric series. Having specified such states we then
build the Bethe states and also identify the spectrum of the model
for generic values of the anisotropy parameter $q$.

\end{abstract}

\vfill

\baselineskip=16pt

\end{titlepage}

\section{Introduction}

Much interest has been focused during the past years on integrable
models with boundaries \cite{cherednik, sklyanin}, further invigorated by advances on
problems related to condensed matter physics and statistical
mechanics, with immediate applications and possible experimental
realizations (see e.g. \cite{ftw, af}), and also by studies on the
understanding of D-branes by means of the boundary conformal field
theory \cite{cardy, polch}.

The main objective of the present work is the study of the open
XXZ model with integrable non-diagonal boundaries for the spin $s$
representation. The spin $s$ XXZ model \cite{kures} with periodic boundary conditions was investigated in \cite{res}, whereas studies concerning the model with toroidal and generic twisted boundary conditions were presented in \cite{alma, batc}. Until recently, the formulation of the Bethe ansatz equations for open spin chains with non-diagonal boundaries
was an open problem. The main difficulty arising when trying to
obtain the transfer matrix eigenvalues and the corresponding Bethe
ansatz equations, by standard algebraic or analytical Bethe ansatz
techniques, is the lack of an obvious reference state
`pseudo--vacuum'. However, the problem was solved rather recently for the spin
${1\over 2}$ XXZ chain by two different methods \cite{nepo, nepo2,
chin}. More precisely, in the approach described in \cite{nepo,
nepo2} such reference state is not a necessary requirement anymore
as long as $q$ is a root of unity. On the other hand in the method
developed in \cite{chin}, which is valid for all values of $q$,
suitable local gauge transformations are implemented, along the
lines described in \cite{baxter, fata}, rendering the derivation of a
suitable reference state possible.

Note that the spectrum of the spin $s$ XXZ model in the presence of
non-diagonal boundaries was derived in \cite{doikou1} using the
method of \cite{nepo}, however only for $q$ root of unity (see also \cite{lima} for spin-1 type chains with diagonal boundaries). Moreover,
in \cite{doikou2} the spectrum, Bethe ansatz equations and Bethe states were
derived for the so called cyclic representation of ${\cal
U}_q(sl_2)$, and for the $q$ harmonic oscillator. The generalized spin $s$ XXX ($q=1$) model with non diagonal boundaries was also investigated within the Bethe ansatz framework in \cite{ann, mart}. Here using the algebraic Bethe ansatz framework
\cite{sklyanin, chin, fata, FT} we are able to specify the spectrum of the spin $s$ XXZ model with non-diagonal boundaries for generic values of $q$, and more importantly deduce the corresponding
eigenstates. It should be stressed that the derivation of the
corresponding eigenstates is of great significance allowing for
instance the calculation of exact correlation functions \cite{korepin, maillet}. The
diagonalization process rests on the identification of an
appropriate reference state upon which all the Bethe states are
built. It is also worth remarking that in the case where the spin
$s$ representation is expressed in terms of difference operators,
the reference state may be derived by solving sets of difference
equations, and it is expressed in terms of $q$-hypergeometric
series.

\section{The open spin $s$ XXZ model}

Before we proceed with the derivation of the spectrum and eigenstates of the spin $s$ XXZ chain with non diagonal boundaries we shall first provide a brief review of the model.
Consider the Lax operator $L(\lambda) \in {\mathbb C}^{2} \otimes {\cal U}_{q}(\widehat{sl_{2}})$ being a solution of the fundamental algebraic relation
\be  R_{12}(\lambda_{1} -\lambda_{2})\ L_{1}(\lambda_{1})\ L_{2}(\lambda_{2})=
L_{2}(\lambda_{2})\ L_{1}(\lambda_{1})\  R_{12}(\lambda_{1} -\lambda_{2}), \label{fund} \ee where
$R$ is the spin ${1\over 2}$ XXZ matrix, solution of the Yang-Baxter equation \cite{baxter, korepin}.
The $L$ operator may be written, with the help of the evaluation homomorphism \cite{jimbo}, which maps
${\cal U}_{q}(\widehat{sl_{2}}) \to {\cal U}_{q}(sl_{2})$, as a $2\times 2$ matrix
\be
L(\lambda) = {1\over 2} \left(
\begin{array}{cc}
e^{\mu \lambda}q^{{1\over 2}} {\mathrm A}-e^{-\mu \lambda}q^{-{1\over 2}} {\mathrm D}    &(q-q^{-1}){\mathrm B}\\
(q-q^{-1}){\mathrm C}    &   e^{\mu \lambda}q^{{1\over 2}} {\mathrm D}-e^{-\mu \lambda}q^{-{1\over 2}} {\mathrm A} \\
\end{array} \right) \label{lax} \ee
${\mathrm A}$, ${\mathrm B}$, ${\mathrm C}$, ${\mathrm D}$ are the generators of ${\cal U}_{q}(sl_{2})$ and they satisfy the well known defining relations:
\be && {\mathrm A}{\mathrm D} ={\mathrm D}{\mathrm A} ={\mathbb I}, ~~~{\mathrm A}{\mathrm C} =q{\mathrm C}{\mathrm A}, ~~~{\mathrm A}{\mathrm B}= q^{-1} {\mathrm B}{\mathrm A}, \non\\ && {\mathrm D}{\mathrm C} =q^{-1}{\mathrm C}{\mathrm D}, ~~~{\mathrm D}{\mathrm B} =q{\mathrm B}{\mathrm D}  ~~~\Big [{\mathrm C},\ {\mathrm B} \Big ] = {{\mathrm A}^{2} -{\mathrm D}^{2} \over q -q^{-1}}. \label{defin} \ee
This algebra admits various realizations, such as the known spin integer or half integer $s$ representation. The $2s+1$ dimensional representation may be also realized by difference operators acting on the space of polynomials of degree 2s (\ref{weyl2}) (see also e.g. \cite{zab}). For the particular case where the generators admit the spin ${1\over 2}$ representation the $L$ operator reduces to the spin ${1\over2}$ XXZ $R$ matrix. More precisely, ${\mathrm A} ={\mathrm D}^{-1} \mapsto q^{\sigma^z},\ {\mathrm B} \mapsto \sigma^-,\ {\mathrm C} \mapsto \sigma^+$, with $\sigma^{z, \pm}$ being the usual $2\times 2$ Pauli matrices.
\\
\\
{\bf (a)} Let us recall the spin $s$ representation of ${\cal U}_{q}(sl_2)$ in more detail. This is a ${\mathrm n}=2{\mathrm s}+1$ dimensional representation, and may
be expressed in terms of ${\mathrm n}\times {\mathrm n}$ matrices
as \be {\mathrm A} = \sum_{k=1}^{{\mathrm n}} q^{\alpha_{k}}\ e_{kk}, ~~~~~{\mathrm C}
=\sum_{k=1}^{{\mathrm n}-1} \tilde C_{k}\ e_{k\ k+1},
~~~~~{\mathrm B}=\sum_{k=1}^{{\mathrm n}-1} \tilde C_{k}\ e_{k+1\ k}
\label{spins} \ee where we define the matrix elements
$(e_{ij})_{kl} = \delta_{ik}\ \delta_{jl}$ and \be \alpha_{k} =
{{\mathrm n}+1 \over 2} - k, ~~~~~\tilde {\mathrm C}_{k} = \sqrt{[k]_q
[{\mathrm n}-k]_q}, ~~~~~[k]_q = {q^k - q^{-k} \over q -q^{-1}}.
\ee
\\
{\bf (b)} Another realization of the spin $s$ representation,
equivalent to the latter one (\ref{spins}), may be given in terms of
the Heisenberg algebra ---essentially in terms of difference
operators. Let  ${\mathrm X}$ and ${\mathrm Z}$ be operators
satisfying the following commutation relation \be  {\mathrm X}\
{\mathrm Z}= q\ {\mathrm Z}\ {\mathrm X},~~~~q = e^{i \mu}.
\label{weyl} \ee It is clear that ${\mathrm X}$ and ${\mathrm Z}$
can be realized in terms of the Heisenberg algebra elements i.e.,
\be {\mathrm X}=\pm e^{\hat p}, ~~{\mathrm Z}=\pm e^{\hat q},
~~\mbox{and} ~~[\pt,\qt]=i \mu\,, \label{weyl2} \ee then the
generators ${\mathrm A}$, ${\mathrm B}$, ${\mathrm C}$, ${\mathrm
D}$ may be expressed as: \be {\mathrm A} =q^{-s}{\mathrm X},
~~~{\mathrm D} = q^{s} {\mathrm X}^{- 1}, ~~~{\mathrm B} =
-{{\mathrm Z}^{-1} \over q-q^{- 1}}({\mathrm X}^{- 1}- {\mathrm
X}),~~~{\mathrm C}= {{\mathrm Z} \over q-q^{- 1}}( q^{2s}{\mathrm
X}^{- 1}- q^{-2s}{\mathrm X}). \non\\ \label{difs} \ee Finally in
the `space' representation $|\textsf{x}\rangle\in{\cal V}^{(R)}$ of
the Heisenberg algebra one defines \be
\pt|\textsf{x}\rangle=i\mu{\partial \over \partial \textsf{x}}
|\textsf{x}\rangle\ ,\qquad
\qt|\textsf{x}\rangle=\textsf{x}|\textsf{x}\rangle. \label{rep} \ee
then it is clear that for any function of
$\textsf{z}=e^{\textsf{x}}$  \be {\mathrm X}^{\pm 1}\ F(\textsf{z})
= F(q^{\pm 1} \textsf{z}). \ee

Recall that our objective here is to study the spin $s$ XXZ chain with non diagonal boundaries.
As is well known to build the model one first needs to construct the open transfer matrix, defined as \cite{sklyanin}
\be &&t(\lambda) = \tr_{0}  \Big \{ K^{+}(\lambda)\  T(\lambda)\
K^{-}(\lambda)\ \hat T(\lambda)\Big \}\ = \tr_{0} \Big \{ K^{+}(\lambda)\
{\cal T}(\lambda)\Big \} ~~~\mbox{where} \non\\
&& T(\lambda) = L_{01}(\lambda)\ \ldots
L_{0N}(\lambda),\ ~~~~~~\hat T(\lambda) = L_{0N}(\lambda) \ldots L_{01}(\lambda)  \label{transfer2}  \ee $L$ is given by (\ref{lax}) and also satisfies $L(\lambda)\ L(-\lambda)  = 2\sinh \mu(\lambda +is +{i\over 2})\sinh \mu (-\lambda +is +{i\over 2})$.
$K^{\pm}$ are solutions of the reflection equation \cite{cherednik}: \be R_{12}(\lambda_1 -\lambda_2)\ K_1(\lambda_1)\ R_{21}(\lambda_1 +\lambda_2)\ K_2(\lambda_2) =K_2(\lambda_2)\ R_{12}(\lambda_1+\lambda_2)\ K_1(\lambda_1)\ R_{21}(\lambda_1-\lambda_2). \label{re} \ee
The general solution is a $2\times 2$ matrix with entries \cite{GZ, DEVGR}:
\be && K_{11}(\lambda)=\sinh\mu(-\lambda+i\xi),
 \qquad
K_{22}(\lambda)=\sinh\mu(\lambda+i\xi) \non\\ &&  K_{12}(\lambda)=\kappa q^{\theta} \sinh(2\mu\lambda), \qquad K_{21}(\lambda)=\kappa q^{-\theta} \sinh(2\mu\lambda). \label{def}\ \ee
Here we shall consider $K^{+} = K(-\lambda -i;\xi^+, \kappa^+, \theta^+)$, and  $K^-(\lambda)= K(\lambda ;\xi^-, \kappa^-, \theta^-)$. Using the fact that ${\cal T}$ satisfies the reflection equation one can show the transfer matrix (\ref{transfer2}) provides a family of commuting operators \cite{sklyanin}: $~[t(\lambda),\ t(\lambda')]=0$.

\section{Bethe states and spectrum}

We can now come to our main aim which is the derivation of the spectrum and Bethe states for the generic open transfer matrix (\ref{transfer2}).
We shall apply in what follows the approach developed in \cite{chin} in order to deduce the eigenstates, the spectrum, and the Bethe ansatz equations of the spin $s$ XXZ chain with non-diagonal boundaries. Define the gauge transformed
$L-$operators
\begin{eqnarray}
{\overline L}_{n}(m|\lambda) &=& \bar M_{m_{n-1}}^{-1}(\lambda)
L_{n}(\lambda)\bar M_{m_{n}}(\lambda)\equiv \left(
        \begin{array}{cc}
 \tilde \alpha_{n} &\tilde \beta_{n} \\
 \tilde  \gamma_{n}  & \tilde  \delta_{n}\\
\end{array}\right)\ ,\non\\ S_{n}(m|\lambda) &=&
M_{m_{n}}^{-1}(-\lambda) L^{-1}_{n}(-\lambda)
M_{m_{n-1}}(-\lambda)\equiv \left(
        \begin{array}{cc}
 \tilde \alpha'_{n} &\tilde \beta'_{n} \\
  \tilde \gamma'_{n}  & \tilde  \delta'_{n}\\
\end{array}\right)\,,~~~m_{n} =m +\sum_{k=1}^{n}g_{k} \label{gauget}\end{eqnarray}
where the local gauge transformations are defined as (for a more detailed description of the method see \cite{chin})
\begin{equation}
M_{n}(\lambda) =  \left(
        \begin{array}{cc}
 e^{-\mu \lambda}x_n &e^{-\mu\lambda}y_n  \\
  1  & 1\\
\end{array}\right),\
~~\bar M_{n}(\lambda) =  \left(
        \begin{array}{cc}
 e^{-\mu \lambda}x_{n+1} &e^{-\mu\lambda}y_{n-1}  \\
  1  & 1\\
\end{array}\right)\label{matrix2} \end{equation}  $x_{n}=-ie^{-i\mu(\beta+\gamma)}e^{- i\mu n} \ , ~y_{n}=-ie^{-i\mu(\beta-\gamma)}e^{i\mu
n}$, and $\beta,\ \gamma$ are $\lambda$ independent ${\mathbb C}$ numbers.
Note that $g_{n}$ depends on the choice of representation and it will be specified later on. For the spin ${1\over 2}$ representation in particular $g=1$ \cite{chin} (for a relevant discussion see also \cite{doikou2}).
The transfer matrix (\ref{transfer2}) can be rewritten with the help of the aforementioned gauge transformations (\ref{gauget}) as (see also Appendix and \cite{chin, doikou2})
\begin{equation}
t(\lambda) = tr_{0} \Big \{ \tilde K^{+}(\lambda)\ \tilde {\cal T}(\lambda) \Big \}
\,, \label{gtransfer} \end{equation}
where $\tilde K^{\pm}$, $~\tilde {\cal T}$ are the gauge transformed matrices (see Appendix).

To diagonalize the transfer matrix (\ref{transfer2}) one first needs a suitable pseudo-vacuum state of the general form:
\be \Omega^{(m)} = \otimes_{n=1}^{N} \varpi_{n}^{(m)}  \label{pseudo1} \ee
$\varpi_{n}^{(m)}$ is the local pseudo-vacuum annihilated by the action of the lower left elements of the transformed $L$ matrices (\ref{gauget}), i.e.
\be \tilde \gamma_{n},~\tilde \gamma'_{n}\ \varpi_{n}^{(m)}=0. \
\label{action} \ee
From the action of $\tilde \gamma_{n}$ on the local pseudo-vacuum the following constraint is obtained
\be \Big [-x_{m_{n}+1}\ (e^{\mu \lambda} q^{{1\over 2}}{\mathrm A}_{n} - e^{-\mu \lambda} q^{{1\over 2}} {\mathrm D}_{n}) +
x_{m_{n-1}+1}\ (e^{\mu \lambda} q^{{1\over 2}}{\mathrm D}_{n} - e^{-\mu \lambda} q^{{1\over 2}} {\mathrm A}_{n}) \non\\ + e^{-\mu \lambda}x_{m_{n}+1}\ x_{m_{n-1}+1}\ (q-q^{-1})\ {\mathrm C}_{n} - e^{\mu \lambda}(q-q^{-1})\ {\mathrm B}_{n}
\Big ] \varpi^{(m)}_{n} =0, \label{con1} \ee a similar constraint arises from the action of $\tilde \gamma'_{n}$ on the local pseudo-vacuum. Notice that the transformed non-diagonal elements of (\ref{k+}), (\ref{k-}) acting on the pseudo-vacuum state (\ref{pseudo1}) must satisfy:
\be &&\tilde K_{2}^{+}(m^{0}|\lambda) =\tilde
K_{3}^{+}(m^{0}|\lambda) = 0\,, ~~~~~\tilde K_{3}^{-}(m_{0}|\lambda) = 0\ \label{cond2} \ee
where the integers $m^{0}$ and $m_{0}$ are associated  to the left and right boundaries respectively. Solving the latter constraints (\ref{cond2}) one can fix the relations between the left and right boundary parameters (see also \cite{nepo, chin, doikou2}). Note however that the problem was solved in \cite{nepo2} in the more general case with no constraints between left and right boundary parameters, however only for $q$ root of unity.

We may now specify the exact reference state for the spin $s$ XXZ model in both realizations (\ref{spins}) and (\ref{difs}).
\\
\\
${\bf  (a)}$ We associate each site of the chain with  the spin $s_{n}$ representation and we
define the local pseudo-vacuum state as
\be \varpi_{n}^{(m)} = \sum_{i=1}^{{\mathrm j}_{n}} w_{l}^{(m, n)}\ f^{(n)}_{l}
\label{pseudo} \ee
$f^{(n)}_{l}$ is the ${\mathrm j}_{n}=2s_{n}+1$ column vector with zeroes everywhere apart from the $l^{th}$ position.

Solving the constraint (\ref{con1}) on the local pseudo-vacuum one can specify the value of ${\bf g_{n} = {\mathrm j}_{n}-1}$ appearing in the local gauge transformations, and we also acquire recursion relations among the coefficients $w_{l}^{(m,n)}$, which
read as:
\be w_{l}^{(m,n)} = {q^{{{\mathrm j}_{n}\over 2} -1}\ (q^{{\mathrm j}_{n}-l+1} - q^{-{\mathrm j}_{n}+l-1}) \over (q-q^{-1})\ \tilde C_{l-1}\ x_{m_{n-1} +1}}\ w_{l-1}^{(m,n)} \label{rec1} \ee with  normalization $w_{1}^{(m,n)} =1$. The solution of the later recursion formula
provides the exact form of the coefficients:
\be w_{l}^{(m,n)} = \Big ( {q^{({{\mathrm j}_{n}\over 2} -1)} \over  x_{m_{n-1} +1}}\Big )^{l-1}
\prod_{k=1}^{l-1}{ [{\mathrm j}_{n} -k]_{q}\over \tilde C_{k}}.\label{so1} \ee
\\
{\bf (b)} Let us now consider the difference realization of the spin $s$ representation (\ref{difs}). In this case the pseudo-vacuum will be expressed as a polynomial of the local spin parameter associated to the $n^{th}$ site of the chain $\textsf{z}^{(n)}$
\be \varpi_{n}^{(m)}= \phi^{(m)}(\textsf{z}^{(n)}) = \sum_{l=1}^{{\mathrm j}_{n}} w_{l}^{(m,n)}\ (\textsf{z}^{(n)})^{l-1}. \ee As in the previous case the aim is to determine the coefficients of the polynomial. The solution of the constraint (\ref{con1}) this time leads to sets of difference equations for the spin variables $\textsf{z}^{(n)}$, namely
\be \Big (\textsf{z}^{(n)} x_{m_{n}+1}\ q^{-{{\mathrm j}_{n}\over 2}+1} +1\Big )\ \phi^{(m)}(q \textsf{z}^{(n)}) -\Big (\textsf{z}^{(n)} x_{m_{n-1}+1 }\ q^{{{\mathrm j}_{n}\over 2}} +1\Big )\ \phi^{(m)}(q^{-1} \textsf{z}^{(n)})=0  . \label{be1} \ee
It is worth stressing that difference equations of the type (\ref{be1}) occur also in the study of the Azbel-Hofstadter problem (see e.g. \cite{wiza, faka}).
Although for our purposes here relations (\ref{be1})
are sufficient we derive, for the sake of completeness, the explicit form
of the polynomial. In any case, the derivation of the pseudo-vacuum
is of great significance per se leading to the explicit from of
the Bethe states as we shall see. In addition we wish to stress
the connection of these polynomials with the $q$-hypergeometric
series. Let us illuminate this point a bit further. Equation
(\ref{be1}) also provides recursion relations among
$w_{l}^{(m,n)}$, whose solution gives the coefficients:
\be w_{l}^{(m,n)} =  (x_{m_{n-1}+1}q^{-{{\mathrm j}_{n} \over 2}+1})^{l-1}\prod_{k=1}^{l-1}{ [{\mathrm j}_{n}-k]_{q}\over [k]_{q}}. \label{rec3} \ee It will be instructive at this stage to express the polynomial $\phi^{(m)}$ in terms of $q$-hypergeometric series (see e.g. \cite{cha} and references therein).
For that purpose it is convenient to rescale the local spin variables as: $\ \bar \textsf{z}^{(n)} = -q^{{{\mathrm j}_{n}\over 2}+1}x_{m_{n-1}+1} \textsf{z}^{(n)}$,\\
then it follows form (\ref{rec3}) that the new coefficients of the polynomial, written in terms of
$\bar \textsf{z}^{(n)}$, are given by
\be \bar w_{l}^{(m,n)} =  \prod_{k=0}^{l-2}{1-q^{-2{\mathrm j}_{n} +2}\ q^{2k} \over 1-q^{2}\ q^{2k}}. \label{rec32} \ee
Let us also introduce some useful notation. Set $\tilde q = q^{2}$, and also define the $q$-shifted factorials as,
\be (a;q)_{n} =\prod_{k=0}^{n-1} (1-aq^{k}), ~~~\mbox{and} ~~~(a_{1}, a_{2}, \ldots, a_{k};q)_{n} = \prod_{j=1}^{k}(a_{j};q)_{n}. \ee Now we may define the so called $q$-hypergeometric functions in the following manner \be _{s+1}\Phi_{s}(a_{1}, a_{2}, \ldots a_{s+1};b_{1}, \ldots b_{s};q,z) = \sum_{k=0}^{\infty} {(a_{1}, \ldots a_{s+1};q)_{k} \over (b_{1}, \ldots, b_{s},q ;q)_{k} } z^{k}, \ee it is thus clear that the polynomial (\ref{rec3}) may be written as a $q$-hypergeometric series \be \phi^{(m)}(\bar \textsf{z}^{(n)}) = \sum_{k=0}^{{\mathrm j}_{n}-1} {(\tilde q^{-{\mathrm j}_{n}+1}; \tilde q)_{k} \over (\tilde q ;\tilde q)_{k} }(\bar \textsf{z}^{(n)})^{k}= _1\Phi_0
(\tilde q^{-{\mathrm j}_{n}+1}; \tilde q, \bar {\textsf z}^{(n)}). \label{qq1} \ee This series obviously terminates because of the presence of the $q^{-{\mathrm j}_{n}+1}$ term in the $q$-shifted factorial appearing in the numerator of (\ref{rec3}).

We have specified the exact form of the reference state for both realizations of the spin $s$ XXZ model (\ref{spins}), (\ref{difs}). To derive the spectrum of the transfer matrix it is also necessary to consider the actions of the transformed diagonal elements on the pseudo-vacuum. Indeed it is relatively easy to show by simply using the transformations (\ref{gauget}), and the recursion relations (\ref{rec3}) that the action of the diagonal elements on the pseudo-vacua, derived above, takes the general form:
\be && \tilde \alpha_{n}^{(m)}\ \varpi^{(m)}_{n} =  a_{n}(\lambda)\ \varpi^{(m+1)}_{n}, \non\\ &&\tilde \delta_{n}^{(m)}\ \varpi^{(m)}_{n} =  {x_{m_{n}+1} -y_{m_{n}-1} \over x_{m_{n-1}+1} -y_{m_{n-1}-1}}\ b_{n}(\lambda)\ \varpi^{(m-1)}_{n} \non\\ && \tilde \alpha_{n}^{'(m)}\ \varpi^{(m)}_{n} = a'_{n}(\lambda)\ \varpi^{(m-1)}_{n}, \non\\ && \tilde \delta_{n}^{('m)}\ \varpi^{(m)}_{n} =   {x_{m_{n-1}} -y_{m_{n-1}} \over x_{m_{n}} -y_{m_{n}}}\
b'_{n}(\lambda)\ \varpi^{(m+1)}_{n} \label{diag3} \ee
where the values of $a_{n}$, $\ b_{n}$, $\ a'_{n}$, $\ b'_{n}$
for each realization are given below
\be  {\bf (a)} && a_{n}(\lambda) = q^{-{\mathrm j}_{n} +1}\sinh \mu (\lambda +{\mathrm j}_{n}{i \over 2}), ~~~~~b_{n}(\lambda) =  q^{{\mathrm j}_{n} -1} \sinh \mu (\lambda -{\mathrm j}_{n}{i \over 2}+i) \non\\  ~~~&& a'_{n}(\lambda) =  q^{{\mathrm j}_{n} -1}\sinh \mu (\lambda +{\mathrm j}_{n}{i \over 2}), ~~~~~b'_{n}(\lambda) =  q^{-{\mathrm j}_{n} +1} \sinh \mu (\lambda -{\mathrm j}_{n}{i \over 2}+i) \non\\ {\bf (b)} && a_{n}(\lambda) = \sinh \mu (\lambda +{\mathrm j}_{n}{i \over 2}), ~~~~~b_{n}(\lambda) =  \sinh \mu (\lambda -{\mathrm j}_{n}{i \over 2}+i) \non\\  ~~~&& a'_{n}(\lambda) =  \sinh \mu (\lambda +{\mathrm j}_{n}{i \over 2}), ~~~~~b'_{n}(\lambda) =   \sinh \mu (\lambda -{\mathrm j}_{n}{i \over 2}+i). \label{diagg}\ee
Our aim now is to solve the following eigenvalue problem
\begin{equation}
t(\lambda)\ \Psi = \Lambda(\lambda)\ \Psi ~~~~~\mbox{where} ~~~~~\Psi  = {\cal B}_{m^{0}-2}(\lambda_1) \ldots {\cal
B}_{m^{0} -2M} (\lambda_M)\ \Omega^{(m)} \
\label{eproblem} \end{equation}
$\Psi$ is of course the general Bethe ansatz state.

The transfer matrix eigenvalues may be deduced by virtue of
algebraic relations emerging from the reflections equation, which
omitted here for brevity (see e.g \cite{chin, doikou2}). The
exchange relations, arising from the reflection equation, are
identical to the ones presented in \cite{chin}. The action
of the diagonal elements on the pseudo-vacuum is naturally
modified, depending explicitly on the choice of the
representation. It may be shown that the state $\Psi$ is indeed an
eigenstate of the transfer matrix if we impose $m\equiv m^0-2M$,
and then it follows that $~m_0=
N g+m^0-2 M$. Let us also define \begin{eqnarray} && {\mathrm f}_{n}(\lambda) = a_n(\lambda) a'_n(\lambda), ~~~~~{\mathrm h}_n(\lambda) = b_n(\lambda)\ b'_n(\lambda) \non\\
&& K_{1}^{+}(m|\lambda) = \tilde K_{1}^{+}(m|\lambda) + {\sinh\mu
(i\gamma+m+2\lambda +i) \sinh i\mu \over  \sinh i \mu
(1+m+\gamma)\sinh\mu (2\lambda +i)}\tilde K_{4}^{+}(m|\lambda)\ , \non\\
&& K_{4}^{+}(m|\lambda) = {\sinh i\mu(m+\gamma) \sinh i\mu \over
\sinh i \mu (1+m+\gamma)\sinh\mu (2\lambda +i)}\tilde
K_{4}^{+}(m|\lambda)\ , \label{k1}
\end{eqnarray} $\tilde K_{1,4}^{\pm}$ are given in the Appendix. Finally the spectrum may be written as
\begin{eqnarray}
&&\Lambda(\lambda) = \Big (
K_{1}^{+}(m^0|\lambda)K_{1}^{-}(m_{0}|\lambda)
\prod_{n=1}^{N}{\mathrm f}_{n}(\lambda)\ \prod_{j=1}^{M} {\sinh
\mu(\lambda +\lambda_{j})\ \sinh \mu(\lambda -\lambda_{j}-i) \over
\sinh \mu(\lambda +\lambda_{j}+i)\ \sinh
\mu(\lambda -\lambda_{j})} \non\\
&+& K_{4}^{+}(m^0|\lambda)K_{4}^{-}(m_{0}|\lambda)
\prod_{n=1}^{N}{\mathrm h}_{n}(\lambda)\ \prod_{j=1}^{M} {\sinh
\mu(\lambda +\lambda_{j}+2i)\ \sinh \mu(\lambda - \lambda_{j}+i)
\over \sinh \mu(\lambda +\lambda_{j}+i)\ \sinh \mu(\lambda
-\lambda_{j})} \Big ) \label{ll}
\end{eqnarray}
provided that certain unwanted terms appearing in the eigenvalue
expression are vanishing. This is achieved as long as
$\lambda_{i}$'s satisfy the Bethe ansatz equations, which are
written below in the familiar form: \be {K_{1}^{+}(m^0|\lambda_i -
{i \over 2})\over  K_{4}^{+}(m^0|\lambda_i - {i \over 2})} {
K_{1}^{-}(m_{0}|\lambda_i - {i \over 2}) \over
K_{4}^{-}(m_{0}|\lambda_i - {i \over 2})  }\ \prod_{n=1}^{N}{
{\mathrm f}_{n}(\lambda_i-{i\over 2}) \over {\mathrm
h}_{n}(\lambda_i-{i\over 2})}  =-\prod_{j=1}^{M} {\sinh\mu
(\lambda_i -\lambda_j +i) \over \sinh\mu (\lambda_{i} -\lambda_j
-i)}\ {\sinh\mu (\lambda_i+ \lambda_j +i) \over \sinh\mu
(\lambda_i+ \lambda_j -i)} \non\\ \label{BAE} \ee $K^{\pm}_{1,4},\ {\mathrm f}_n,\ {\mathrm h}_n$ are defined in (\ref{k1}), and this completes the
derivation of the spectrum and Bethe states of the spin $s$ XXZ
chain with non-diagonal boundaries.

\section{Discussion}

Let us briefly recall the main findings of the present
investigation. We have been able to derive the spectrum and Bethe
ansatz for the open spin $s$ XXZ chain with special non-diagonal
boundaries and for generic values of the anisotropy parameter. Note
that the spectrum (\ref{ll}), and Bethe ansatz equations (\ref{BAE})
for the spin $s$ XXZ model with non-diagonal boundaries are similar
with the ones found in \cite{doikou1}, although the investigation in
\cite{doikou1} was restricted for $q$ root of unity. More
importantly in the present study we were able to specify the Bethe
states of the model by means of suitable local gauge
transformations. The first important step was the derivation of an
appropriate reference arising as a solution of certain difference
equations (\ref{rec1}), (\ref{be1}). Then within the algebraic Bethe
ansatz framework we built the corresponding Bethe states and derived
the spectrum and Bethe ansatz equations.
\\
\\
\noindent{\bf Acknowledgments:} This work was supported by INFN, and the European Network `EUCLID'; `Integrable
models and applications: from strings to condensed matter',
contract number HPRN--CT--2002--00325.

\appendix

\section{Appendix}

The local gauge transformations may be expressed as:
\begin{equation}
M_{n}(\lambda) = \Big (X_{n}(\lambda), ~~Y_{n}(\lambda) \Big),
~~\bar M_{n}(\lambda) = \Big (X_{n+1}(\lambda), ~~Y_{n-1}(\lambda)
\Big) \label{matrix2a} \end{equation} and \begin{equation}
X_{n}(\lambda)= \left (
\begin{array}{c}
  e^{-\mu \lambda}x_{n}\\
  1 \\
\end{array}
\right)\,, ~~ \ \ \ Y_{n}(\lambda)= \left (
\begin{array}{c}
  e^{-\mu \lambda}y_{n}\\
  1 \\
\end{array}
\right)\,, \qquad x_{n}=x_0e^{- i\mu n} \ , \quad y_{n}=y_0e^{i\mu
n}\ .\label{col}
\end{equation}
It is also convenient to introduce the matrices
\be M_{n}^{-1}(\lambda) = \left ( \begin{array}{c}
\bar Y_{n}(\lambda)\\
\bar X_{n}(\lambda)\\
\end{array}
\right)\,,~~\bar M_{n}^{-1}(\lambda) = \left ( \begin{array}{c}
  \tilde Y_{n-1}(\lambda)\\
  \tilde X_{n+1}(\lambda)\\
\end{array}\right)\ ,
\ee with
\begin{eqnarray} \bar X_{n}(\lambda)= {1\over x_{n} -y_{n}}
\Big (-e^{\mu \lambda}, ~~x_{n} \Big  ), ~~ \bar Y_{n}(\lambda)=
{1\over x_{n} -y_{n}}\Big (e^{\mu \lambda}, ~~-y_{n}\Big  )\ ,
 \non\\ \tilde X_{n}(\lambda)= {1\over x_{n} -y_{n-2}}
\Big (-e^{\mu \lambda}, ~~x_{n} \Big  ), ~~ \tilde Y_{n}(\lambda)=
{1\over x_{n+2} -y_{n}}\Big (e^{\mu \lambda},
 ~~-y_{n}\Big  )\ . \label{col2} \end{eqnarray}
The transformed $K^{\pm}$ and ${\cal T}$  matrices are expressed as:
\begin{equation}
\tilde K^{+}(m|\lambda) =  \left(
        \begin{array}{cc}
 \tilde K_{1}^{+}(m|\lambda) &\tilde K_{2}^{+}(m|\lambda)  \\
  \tilde K_{3}^{+}(m|\lambda)  &\tilde K_{4}^{+}(m|\lambda)  \\
\end{array}\right)=  \left(
\begin{array}{cc}
\bar Y_{m}(-\lambda)K^{+}(\lambda) X_{m}(\lambda) &\bar Y_{m}(-\lambda)K^{+}(\lambda)Y_{m-2}(\lambda)  \\
\bar X_{m}(-\lambda)K^{+}(\lambda) X_{m+2}(\lambda)  &\bar X_{m}(-\lambda)K^{+}(\lambda)Y_{m}(\lambda)  \\
\end{array}\right)\ \label{k+}
\end{equation} and \begin{equation}
\tilde {\cal T}(\lambda) =  \left(
        \begin{array}{cc}
  {\cal A}_{m}(\lambda) &{\cal B}_{m}(\lambda)  \\
   {\cal C}_{m}(\lambda)  & {\cal D}_{m}(\lambda)  \\
\end{array}\right)\ =\left(
\begin{array}{cc}
\tilde Y_{m-2}(\lambda) {\cal T}(\lambda)X_{m}(-\lambda) &\tilde Y_{m}(\lambda){\cal T}(\lambda) Y_{m}(-\lambda)  \\
\tilde X_{m}(\lambda){\cal T}(\lambda)X_{m}(-\lambda)  &\tilde X_{m+2}(\lambda){\cal T}(\lambda)Y_{m}(-\lambda)  \\
\end{array}\right)\,.\label{U}
\end{equation}
Similarly to \cite{chin}, one defines the transformed $K^{-}$
matrix as
\begin{equation} \tilde K^{-}(l|\lambda)=
\left(\begin{array}{cc}
\tilde K_{1}^{-}(l|\lambda) &\tilde K_{2}^{-}(l|\lambda)  \\
\tilde K_{3}^{-}(l|\lambda) &\tilde K_{4}^{-}(l|\lambda)  \\
\end{array} \right)=
\left(\begin{array}{cc}
 \tilde Y_{l-2}(\lambda)K^{-}(\lambda) X_{l}(-\lambda) &\tilde Y_{l}(\lambda)K^{-}(\lambda)Y_{l}(-\lambda)  \\
\tilde X_{l}(\lambda)K^{-}(\lambda) X_{l}(-\lambda)  &\tilde X_{l+2}(\lambda)K^{-}(\lambda)Y_{l}(-\lambda)  \\
\end{array}\right)\ \label{k-} \end{equation} with $l=m+\sum_{n=1}^{N}g_{n}$.

\end{document}